\documentclass[a4paper,10pt]{article}
\title{Exceptional Polynomials and SUSY Quantum Mechanics}
\author{K. V. S. Shiv Chaitanya$^1$, S. Sree Ranjani$^2$, Prasanta K. Panigrahi,$^3$\\ R Radhakrishnan$^4$ and V. Srinivasan$^4$\\
$^1$Institute of Mathematical Sciences, Taramani, Chennai, India, 600113.\\$^2$School of Physics, University of Hyderabad, Hyderabad, India, 500 046.\\
$^3$Indian Institute of Science Education and Research (IISER)\\ Kolkata, Mohanpur Campus, Nadia, India, 714252.\\
$^4$Department of Theoretical Physics, University of Madras,\\  Guindy Campus, Chennai, India, 600025.\footnote{$^1$chaitanya@imsc.res.in, $^3$pprasanta@iiserkol.ac.in}}

\begin{document}

\maketitle
\begin{abstract}
We show that the existence of exceptional polynomials leads to the presence of non-trivial supersymmetry. The existence of these polynomials reveals several distinct isospectral potentials  for the Schr\"odinger equation.
All Schr\"odinger equations having Laguerre and Jacobi polynomials as their solutions, have non-trivial supersymmetric partners with corresponding exceptional polynomials as solutions.  
\end{abstract}

\section{Introduction}

The celebrated theorem of Bochner \cite{boc} states that if an infinite sequence of 
polynomials $P_n(x)= y_n  \,\,\,( n=0,1,2\cdots \infty)$ satisfies a second order eigenvalue equation
of the form,
\begin{equation}
 p(x) y_n'' + q(x) y_n' + r(x) y_n=\lambda_n y_n,      \label{sl}
\end{equation}
then $p(x),\, q(x)$ and $r(x)$ are polynomials of order $2,\, 1$ and $0$ respectively. 
This theorem has been extended to $q$-difference equations by Askey \cite{ask}. The polynomials $y_n$ form a complete set with respect to a positive measure. Till recently it was thought that only the classical orthogonal polynomial systems (OPS) like, the Hermite, Laguerre and the Jacobi satisfy eq.(\ref{sl}).  Note that $y_0$ is a constant in all these systems. It was shown \cite{kam,kam1} that it is possible to construct an OPS, which starts with $y_n, \,( n=1,2 \cdots)$  by a suitable modification of the weight function which forms a complete set. The newly constructed OPS was shown to satisfy a Sturm-Liouville equation of the form given in eq.(\ref{sl}), where $ r(x)$ is not a constant. Only the Laguerre and the Jacobi polynomials were shown to allow such extensions, which are known as exceptional polynomials.

\noindent
One can construct the exceptional $X_1$-Laguerre Polynomials $\mathcal{L} _n^k(x),\, k>0$, using the Gram-Schmidt procedure from the sequence \cite{kam,kam1},
\begin{equation}
 v_1=x+k+1 ;\; v_i=(x+k)^i , i\ge 2,
\end{equation}
using the weight function 
\begin{equation}
 \hat{W}_k(x)= \frac {x^k e ^{-x}} {(x+k)^2},  \label{Wlag}
\end{equation}
defined in the interval $x \in (0,\infty)$ and the scalar product 
\begin{equation}
 (f,g)_k= \int _0 ^{\infty} dx \hat{W}_k(x) f(x)g(x). 
\end{equation}
The weight function for the normal Laguerre polynomial $W_k(x)=x^k e ^{-x}$, is multiplied by suitable factors 
such that one obtains a new $\hat{W}_k(x)$ such that one  can construct the new OPS which does not have the zero degree polynomial.
The exceptional $X_1$-Laguerre differential equation is 
\begin{equation}
 T_k (y)  = \lambda y, \label{el}
\end{equation}
where $\lambda=n-1$ and 
\begin{equation}
T_k(y) = -x y'' +\left ( \frac{x-k}{x+k} \right ) [(k+x+1)y'-y].  \label{lau}
\end{equation}
For more details we refer the reader to \cite{kam, kam1} and the references therein.  

Similarly the exceptional $X_1$-Jacobi polynomials $\mathcal{P}^{(\alpha,\beta)}_n(x)$, for $\alpha \neq \beta$ and real, are obtained from the sequence
\begin{equation}
 u_1= x-c, \, \, u_i=(x-b)^i,\, i \ge 2,
\end{equation}
where
\begin{equation}
a=\frac{1}{2}(\beta-\alpha); b=\frac{\beta+\alpha}{\beta-\alpha},\;  \textrm{ and }\; c=b+\frac{1}{a}.
\end{equation}
The scalar product is defined in the range $[-1,1]$ with the weight function
\begin{equation}
\hat{W}_{\alpha,\beta}(x)= \frac {(1-x)^{\alpha} (1+x)^{\beta}}{(x-b)^2}.  \label{Wjac}
\end{equation}
As in the case of exceptional Laguerre case, the weight function of this new OPS is a rational extension of the classical Jacobi  weight function, $W_{\alpha,\beta}(x)=(1-x)^{\alpha} (1+x)^{\beta}$.

These exceptional $X_1$-Jacobi polynomials obey the eigenvalue equation 
\begin{equation}
(x^2-1)y''+2a\left ( \frac{1-bx}{b-x}\right) [(x-c)y'-y]= \lambda y,
\label{ej}
\end{equation}
where $\lambda = (n-1)(\alpha+\beta+n)$.

It is well known that the solutions of the Schr\"odinger eigenvalue  problems, with exactly solvable (ES) potentials involve classical OPS. The form of the complete solution is 
$ \sqrt{W(x)}y_n$, where $W(x)$ is the weight function corresponding to the OPS, $y_n  (n=0,1,2\ldots)$,  such that $\int _a ^b W(x) y_n(x)y_m(x) dx = \delta_{mn}$. The weight function is determined from the boundary condition of the given potential. By adding suitable terms to the potential, we can  modify the weight function, so that we obtain exceptional polynomials as solutions to the  Schr\"odinger equation (SE). We make a crucial observation that a quantum mechanical problem which admits classical Laguerre/ Jacobi polynomials as a solutions of the SE, after a modification of the potential will admit exceptional Laguerre/ Jacobi polynomials as solutions having same eigenvalues, with the ground state missing in one of the Hamiltonians.  

\section{Exceptional Polynomials in Quantum Mechanics}

Theorem : By adding an extra term $V_e(x)$ to the Laguerre/ Jacobi differential equation and demanding the solutions to be  $g(x)=\frac{f(x)}{(x+m)}$ and $g(x)=\frac{f(x)}{(x-b)}$
for the Laguerre and Jacobi respectively, where $f(x)$ satisfies $X_1$- exceptional differential equation for the Laguerre and Jacobi respectively, $V_e(x,m)$ can be determined uniquely.

Proof : Let  $g(x)=L_{\lambda}^{m} (x)$ satisfy the Laguerre differential equation 
\begin{equation}
  x\frac{d^2}{dx^2}g (x) + (m+1-x) \frac{d}{dx}g (x) +(\lambda -m) g (x) = 0,\label{lague}
\end{equation}
where $\lambda$ is an integer. By adding an extra term $V_e(x,m)$ to the
Laguerre differential equation: \begin{equation}
  x\frac{d^2}{dx^2}g (x) + (m+1-x) \frac{d}{dx}g (x) +(\lambda + V_e(x,m) -m) g (x) = 0, \label{laguet}
\end{equation}
and  setting $g(x)=\frac{f(x)}{(x+m)}$, where $f(x)$
satisfies the $X_1$ exceptional Laguerre differential equation
\begin{equation}
-x f''(x) +\left ( \frac{x-m}{x+m} \right ) [(m+x+1)f'(x)-f(x)]=(n-1)f(x), \label{exlau}
\end{equation}
determines $V_e(x,m)$ to be
\begin{equation}
V_e(x,m)=\frac{1}{(x+m)}-\frac{2m}{(x+m)^2}.  \label{couexr}
\end{equation}
If $g(x)=\frac{f(x)}{(x+m)^j}$ with $f(x)$ satisfying the $X-j$ exceptional differential equation
\begin{equation}
 -x f''(x) +\left ( \frac{x-m}{x+m} \right ) \left[\left((m+x+1)-\frac{2x(j-1)}{x-m}\right)f'(x)-jf(x)\right]= (n-j)f(x),
\end{equation} 
one obtains $V_{e}(x,m)$ to be
\begin{equation}
V_e(x,m)=\frac{j}{(x+m)}-\frac{j(j+1)m}{(x+m)^2}\label{my}
\end{equation}
for the general case.  
Similarly for the Jacobi polynomials, $g(z)=P_n^{(s - \lambda - 1/2, s + \lambda - 1/2)} (z)$, satisfying the Jacobi differential equation
\begin{equation}
 (1-z^2)g''(z)+[\beta-\alpha-(\alpha+\beta+2)z]g'(z)+n(n+\alpha+\beta+1)g(z)=0, \label{jact}
\end{equation} 
adding an extra potential $V_e(z)$ to it gives
\begin{equation}
 (1-z^2)g''(z)+[\beta-\alpha-(\alpha+\beta+2)z]g'(z)+n(n+V_e(z)+\alpha+\beta+1)g(z)=0.\label{jact1}
\end{equation} 
Setting $g(z)=\frac{f(z)}{(z-b)}$ and demanding that $f(z)$ satisfy the $X_1$ exceptional Jacobi differential equation
\begin{equation}
(z^2-1)f''(z)+2a\left ( \frac{1-bz}{b-z}\right) [(z-c)f'(z)-f(z)]= \lambda f(z),
\label{ej1}
\end{equation}
where $\lambda = (n-1)(\alpha+\beta+n)$. This determines $V_{e}(z)$ to be
\begin{equation}
V_{e}(z)=\frac{2}{(z-b)}-\frac{2b}{(z-b)^2}  \label{jac1}.
\end{equation}
This ends the proof of our theorem.

\noindent
Example(1): For 3D oscillator case in natural units the radial equation is solved by
\begin{equation}
 R(\xi)=\xi^{\frac{l}{2}}e^{-\frac{\xi}{2}}K(\xi)
\end{equation}
where $K(\xi)$ satisfies Laguerre differential Equation 
\begin{equation}
  \xi\frac{d^2}{d\xi^2}K (\xi) + (l+\frac{3}{2}-\xi) \frac{d}{d\xi}K (\xi) +(\lambda -3-2l) K (\xi) = 0,\label{osc}
\end{equation}
 with $\xi=r^2$.
where $\lambda=2n+3$, $n=l+2n'$ and $n'=0,1,2....$ and $\lambda=2E$, which gives $E=n+\frac{3}{2}$. Identifying with the eq. (\ref{lague}) one has $m=l+ \frac{1}{2}$ and  $\lambda=n'+l+ \frac{1}{2}$. Then one gets
\begin{equation}
V_{e}^{osc}(\xi,l)=\frac{1}{(\xi+\frac{2l+1}{2})}-\frac{2(\frac{2l+1}{2})}{(\xi+\frac{2l+1}{2})^2}\label{osc1}.  
\end{equation}
Thus one obtains the modified 3D oscillator potential, studied in \cite{qu1}, \cite{od}, by adding $V_{e}^{osc}(\xi,l)$ to the 3D oscillator potential which was. A quantum Hamilton-Jacobi  analysis of the modified oscillator potential has been done in \cite{pkp}.
\noindent
Example(2) As is well known, the radial equation for the Coulomb potential in natural units is 
\begin{equation}
\frac{d^2}{dr ^2}R(r)+\frac{2}{r}\frac{d}{dr}R(r)
+ \left[\frac{\lambda}{r}-\frac {1}{4} - \frac{ l(l+1)}{r ^2} \right] R(r) = 0. 
\label{rad}
\end{equation}
The solution to $R(r)$ is of the form
\begin{equation}
 R(r)=r^le^{-\frac{r}{2}}K(r),
\end{equation}
where $K(r)$ satisfies Laguerre differential equation 
\begin{equation}
  r\frac{d^2}{dr^2}K (r) + (2(l+1)-r) \frac{d}{dr}K (r) +(\lambda -l-1) K (r) = 0.\label{colp}
\end{equation}
 Thus, $K(r)=L_{n'+l}^{2l+1}(r)$
comparing this with eq. (\ref{lague}) one gets $\lambda=n'+l$, $m=2l+1$ and $\lambda=\frac{1}{\sqrt{-2E}}$.
Now, by replacing $\lambda$  by $\lambda +V_e(r,l)$ then the extra term to eq. (\ref{rad}) and eq.(\ref{colp}) takes the form
\begin{equation}
V_{e}(r,l)=\frac{1}{(r+2l+1)}-\frac{2(2l+1)}{(r+2l+1)^2}.  \label{couex}
\end{equation}
Example(3) For the Morse potential \cite{kharebook}, the radial equation is solved by
\begin{eqnarray}
 R(y)=y^{s-n} e^{-\frac{1}{2} y}K(y),  
\end{eqnarray}
where $K(y)$ satisfies Laguerre differential equation 
\begin{equation}
  y\frac{d^2}{dy^2}K (y) + (2(s-n)+1-y) \frac{d}{dy}K (y) +(\lambda -(s-n)) K (y) = 0.\label{mor}
\end{equation}
Here $K(y)=L_n^{2s-2n}(y)$ with $y = {2B \over \alpha} e^{\alpha x}$, and $ s = A/\alpha$ and the eigenvalues,  $\lambda= A^2 - (A - n\alpha)^2$. Then the exceptional part is given by 
\begin{equation}
V_{e}^{Mor}(y,s)=\frac{1}{(y+s-n)}-\frac{2(s-n)}{(y+s-n)^2}\label{mor1}.
\end{equation}
The solution of  the Schr\"odinger  equation with  extra potential $V_e(x)$ added for bound state problems form a complete set. Note that solutions are of the form $\sqrt{W(x)}\mathcal{L}_\lambda^m$ this gives a proof for the completeness of the exceptional polynomials.

It should be noted that all these potentials admitting the Laguerre differential equation (\ref{lague}) can be brought to the same form by a point canonical transformation. They only differ in the constant values of $\lambda$ and $m$. For example, in natural units $V^{osc}=\frac{1}{2}x^2-E$, by making a change of variable $y=x^2$, one gets the Coulomb potential $V^{Coul}=\frac{V^{osc}}{yE}=\frac{1}{2E}-\frac{1}{y}$, without altering the centrifugal term.  Similarly, starting from the Morse potential, $V=A^2+B^2\exp(-2\alpha x)-2B(A+\alpha/2)\exp(-\alpha x)$, by change of variable $r=\exp(\alpha x)$ goes over to the Coulomb potential.

A similar analysis goes through for the Jacobi polynomials. Consider the Scarf potential \cite{qu,qu1}:
\begin{equation}
V(x)= -A^2+(A^2+B^2-A \alpha) \sec ^2 (\alpha x)-B(2A - \alpha) \tan (\alpha x),  \label{sc1}
\end{equation}
whose solution  is
\begin{equation}
\psi(z)= (1 - z)^{(s - \lambda)/2} (1 + z)^{(s + \lambda)/2} 
			P_n^{(s - \lambda - 1/2, s + \lambda - 1/2)} (z),  \label{sc2}
\end{equation}
where $z=\sin (\alpha x)$,  $s = A/\alpha$ and $\lambda = B/\alpha $. Here, $P_n^{(s - \lambda - 1/2, s + \lambda - 1/2)} (z) \equiv y$  are the classical Jacobi polynomials which satisfy the Jacobi differential equation
\begin{equation}
 (1-z^2)y''+[\beta-\alpha-(\alpha+\beta+2)z]y'+n(n+\alpha+\beta+1)y=0. 
\end{equation} 
Adding an extra potential $V_e(z)$ and demanding the solution to be of the form,
\begin{equation}
 \psi(z)= \frac{(1 - z)^{(s - \lambda)/2} (1 + z)^{(s + \lambda)/2}}{z-b} 
 \mathcal{ P}_n^{(s - \lambda - 1/2, s + \lambda - 1/2)} (z),   \label{sc3}
\end{equation}
with $\mathcal {P}_n^{(s - \lambda -1/2, s + \lambda - 1/2)} (z)$ satisfying the $X_1$-Jacobi differential equation eq.(\ref{ej}), 
determines $V_{e}(z)$ to be
\begin{equation}
V_{e}(z)=\frac{A(2A-1)}{2A-1-2B z}-\frac{A(2A-1)^2-4B^2}{(2A-1-2B z)^2}.  \label{jac}
\end{equation}
The Scarf potential in eq. (\ref{sc1}) has the Jacobi polynomials as solutions and all the potentials having Jacobi polynomials as solutions can be mapped  to the Scarf potential by a suitable change of variable. 


\section{Supersymmetric Quantum Mechanics}

The discovery of exceptional polynomials led Quesne to guess the superpotential $\mathcal{W}(x)$, demanding isospectrality \cite{qu1}. 
Supersymmetric quantum mechanics assumes isospectral Hamiltonians. In the previous section, we have constructed the two sets of isospectral Hamiltonians, of which, one Hamiltonian denoted by  $H^+$ with potential $V^+$ has the Laguerre/ Jacobi solution  and another Hamiltonian denoted by  $H^-$ with potential $V^-$ has exceptional Laguerre/ Jacobi solution. Therefore, it is natural ask, ``can the two potentials thus constructed be partner potentials? Can one construct a superpotential, $mahcal{W}(x)$, from which these two can be constructed?'' Here $V^-$ cannot be obtained from $V^+$ through the shape invariant arguments. Below, we answer these questions in the affirmative.

First, we briefly review conventional SUSY and refer the reader to \cite{khareajp} and \cite{kharebook} for more details.
In supersymmetry, the superpotential $\mathcal{W}(x)$ is defined in terms of the intertwining operators $ \hat{A}$ and $\hat{A}^{\dagger}$  as
\begin{equation}
  \hat{A} = \frac{d}{dx} + \mathcal{W}(x), \qquad \hat{A}^{\dagger} = - \frac{d}{dx} + \mathcal{W}(x).
\label{eq:A}
\end{equation}
This allows one to define a pair of factorized Hamiltonians $H^{\pm}$ as
\begin{eqnarray}
   H^{+} &=& 	\hat{A}^{\dagger} \hat{A} 	= - \frac{d^2}{dx^2} + V^{+}(x) - E, \label{vp}\\
  H^{-} &=& 	\hat{A}  {\hat A}^{\dagger} 	= - \frac{d^2}{dx^2} + V^{-}(x) - E, \label{vm}
\end{eqnarray}
where $E$ is the factorization energy. 

The partner potentials $V^{\pm}(x)$ are related to $\mathcal{W}(x)$ by 
\begin{equation}
  V^{\pm}(x) = \mathcal{W}^2(x) \mp \mathcal{W}'(x) + E, \label{gh}
\end{equation}
where  prime denotes differentiation with respect to $x$. The eqs.(\ref{vp}) and (\ref{vm}), imply
\begin{equation}
   H^{+} \hat{A}^{\dagger} = \hat{A}^{\dagger} H^{-}, \qquad \hat{A} H^{+} = H^{-} \hat{A}.\label{df}
\end{equation}
From the above one can see that the operators $\hat{A}$ and $\hat{A}^{\dagger}$ act as a intertwining operator. These operators allows one to go from wave function $\vert\psi^{+}_{\nu} \rangle$ to the wave function $\vert\psi^{-}_{\nu} \rangle$ and vice versa. Our aim is to construct the operator $\hat{A}$ for the isospectral Hamiltonians given in the previous section. We would also like to add that the conventional methods for obtaining the superpotential from the ground state eigenfunction \cite{od, asim} does not work here. In our case,  the wave functions for both the Hamiltonians are known and one can construct the operator $\hat{A}$ from the following relation 
\begin{equation}
\hat{A}H^{+}\vert\psi^{+}_{\nu} \rangle=E_\nu \hat{A}\vert\psi^{+}_{\nu} \rangle
=E_{\nu+1}\vert\psi^{-}_{\nu+1} \rangle=H^{-}\vert\psi^{-}_{\nu+1} \rangle\label{oper}
\end{equation}
using the operator $\hat{\cal{O}}$   \cite{kam, kam1, qu1}, which connects the ordinary Laguerre polynomials to the exceptional Laguerre
polynomials 
\begin{equation}
\hat{\cal{O}} L_{\nu}^{k-1}(x)={\cal L}^{k}_{\nu+1}(x),    \label{34}
\end{equation}
where $\cal{O}=(x+k)(\frac{d}{dx}-1)-1$. The superpotential, ${\cal W}(x)$, can be  obtained by replacing $\frac{d}{dx}$ in  $\hat{A}$ in terms of $\hat \mathcal{O}$. The superpotential ${\cal W}(x)$ determined to be
\begin{equation}
\mathcal{W}(x)= -\frac{l}{x}-\frac{1}{2}-\frac{1}{x+k}.\label{imp}
\end{equation} 
We obtain the intertwining operators 
defined in eq. (\ref{eq:A}), which take one from $\vert\psi^{+}_{\nu} \rangle$ to $\vert\psi^{-}_{\nu} \rangle$ and vice-versa, which involve classical and exceptional Laguerre polynomials. Thus, the form of these intertwining operators is universal. Hence, we prove the existence of exceptional polynomials leads to the presence of non-trivial supersymmetry.

 It should be noted that to find the isospectral potentials, it is not necessary that the Hamiltonians be of the form  $\hat{A}^\dagger\hat{A}$ and $\hat{A}\hat{A}^\dagger$.  It turns out that in the case of 3D oscillator, since the Hamiltonian is positive definite, the supersymmetry machinery goes through and the partner Hamiltonians are of the form $\hat{A}^\dagger\hat{A}$ and $\hat{A}\hat{A}^\dagger$ and  we recover the results of Quesne \cite{qu1}, for appropriate identification of the variables.
By taking $V^{+}_l(x)=\frac{1}{2}x^2+\frac{l(l+1)}{x^2}-E$ and $V^{-}(x)=V_{l-1}^{-}(x)+V_e(x)$
then one  obtains 
\begin{eqnarray}
2\mathcal{W}'(r)=V^{osc}_{e} =V^{+}- V^{-}&=&-\frac{l}{x^2}+\frac{2x^2}{(x^2+k)^2}-\frac{1}{(x^2+k)}.
\end{eqnarray}
From this one gets the superpotential $\mathcal{W}^{\prime}(x)$ for the 3D oscillator. 
However, in the Coulomb problem one has negative eigenvalues. This makes the Hamiltonian negative definite. This can be overcome by taking $-\hat{A}^\dagger\hat{A}$ to be $\hat{B}^\dagger\hat{B}$, which is positive semi-definite.
Or one can always add a zero point energy as done in conventional SUSY, i.e $H_0+\epsilon=\epsilon+E'$, where
$E'$ is the conventional energy of the bound state Hydrogen atom which can be written as $\lambda=\frac{1}{\sqrt{-2E'}}$,  that makes $H_0+\epsilon$ positive semi-definite without changing the wave function. The superpotential of the Coulomb problem is obtained by mapping it to the oscillator problem, with the following change of variable $r=x^2$.  One gets $2\mathcal{W}'(x)$
to be
\begin{eqnarray}
2\mathcal{W}'(x) =\frac{V^{osc}_{e}}{r}&=&-\frac{l}{r^2}+\frac{1}{r}(\frac{1}{(r+k)}-\frac{2k}{(r+k)^2}).
\end{eqnarray}

Similar method will work for ES potentials which have classical Jacobi polynomials as solutions.  Here the operator, which allows us to go from classical to exceptional $X_1$-Jacobi polynomials is given as 
\begin{equation}
 O_j P^{\alpha-1, \beta+1}_n (x)=2(\beta-\alpha)(\beta+n)\mathcal{P}^{\alpha, \beta}_{n+1},
\end{equation}
where $ O_j=[\alpha+\beta -(\beta-\alpha)x]\left( (1+x) \frac{d}{dx}+ \beta +1 \right ) +  (\beta-\alpha)(1+x)$. Taking the wave functions given in Eqs. (\ref{sc2}) and (\ref{sc3}) as $\psi^{+}(x)$ and $\psi^{-}(x) $ respectively, we can construct  $\mathcal{W}(x)$, which allows us to construct the intertwining operators whose form is universal. 
We recover Quesne's results for the Scarf potential constructed in ref \cite{qu1}.

\section{Conclusion}

We have proved a theorem that whenever the Schr\"odinger equation is associated with the Laguerre/ Jacobi polynomials as solutions, one can also construct the exceptional Laguerre/ Jacobi polynomials as solutions for the Schr\"odinger equation by adding an extra term $V_{e}(x)$ to the original potential.
The form of the potential is universal and it depends on the extra terms, $1/(x+k) $ for the Laguerre and $1/(x -b) $ for the Jacobi polynomials. We have also proved this theorem for the general $X^j$ exceptional Laguerre/Jacobi polynomials. We have shown that these two set of Hamiltonians satisfy new kind of supersymmetry called nontrivial supersymmetry.

\section*{Acknowledgments}
Authors thank M. S. Sriram and A. K. Kapoor for stimulating conversations. 
VS thanks the Department of Theoretical Physics, Madras University 
for a visiting professorship during which this work was done. SSR acknowledges the Department of Science and technology, Govt of India (fast track scheme (D. O. No: SR/FTP/PS-13/2009)) for financial support.


\end{document}